# Work function tuning of graphene oxide by using cesium applied to low work function tethers

S. Naghdi[1], A. Várez, X. Chen, J. Navarro, G. Sánchez-Arriaga
*Universidad Carlos III de Madrid, Spain*

J. Fabian-Plaza, G. Meiro, A. Post
*Advanced Thermal Devices, Spain*

**Low Work Function Tethers (LWT) are long conductors with a segment coated with a material that have loosely-bounded electrons. Unlike a standard bare tether equipped with an active electron emitter, LWTs close the electrical circuit (cathodic contact) with the ambient plasma in a passive manner thanks to the thermionic and photoelectric effects. This work presents experimental results on a novel procedure for manufacturing LWT samples. Conductive substrates (aluminum and copper) have been coated with graphene oxide doped with Cesium by using a scalable, simple and cost-effective coating method with an air brush. Both GO and Cs-doped GO (GO/Cs) samples were characterized utilizing X-ray spectroscopy, field emission scanning electron microscopy, four-point probe, and ultraviolet photoelectron spectroscopy (UPS). They showed that GO sheets were successfully doped with Cs atoms, which do not agglomerate on the graphene sheets and do not form particles. The UPS results indicated that by doping GO sheets with Cs, the work function of GO decreases from 4.6 eV to 3.09 eV, thus increasing the applicability of GO as an electron emitter material.**

## Nomenclature

| | | |
|---|---|---|
| *LWT* | = | Low work function tether |
| *eV* | = | Electron volt |
| *GO* | = | Graphene oxide |
| *EDS* | = | Energy-dispersive X-ray spectroscopy |
| *FE-SEM* | = | Field emission electron microscopy |
| *UPS* | = | Ultraviolet photoelectron spectroscopy |

## 1. Introduction

A great variety of terrestrial and space devices need efficient electron emission. Electronic and certain energy conversion devices, hollow cathodes, and thermionic emitters are some illustrative examples. The performance and efficiency of these and many other devices are often controlled by their work function (W), which is defined as the difference between the vacuum level and the Fermi Level and represents the minimum required energy to remove an electron from the material to the vacuum immediately outside the solid surface. Since the electron emission processes are typically very sensitive to the value of W, the development of new stable materials with lower W is a current challenge in material science. An example is the thermionic emission that follows the Richardson-Dushman law and exhibits an exponential dependence with W. For that reason, even a small W-reduction can produce important benefits in terms of performance. Engineering the band structure of the electron emitter materials for tuning the W is an active field of research.

Since the isolation of graphene and the discovering of its extraordinary characteristics, a variety of applications have been proposed to take advantage of this versatile material. Nowadays some of them already found their way into the markets, while there are other promising uses, which developing daily to be suitable enough (cost effective, ecofriendly, comparable to the already exist products...) for industrial scale production. The space sector is not an exception and graphene has been proposed for making energy-store devices [1], loop heat pipes [2], and solar sails [3], among others. As explained below, using graphene as an electron source for thermionic emission devices is another ongoing research project that can open new horizons in space applications. Regarding electrodynamic tethers, there are, at least, two potential applications: the use of graphene in active electron emitters in bare tether configurations [4] and its use in the coating of low work function tethers [5, 6]. For both cases, a material with loosely-bounded electrons is needed.

Graphene properties, including a high electrical and thermal conductivity, ultrahigh electrical mobility, outstanding mechanical properties, ability to withstand high temperatures (as high as 4600 K in the vacuum), and high emission density, indicate that it can be a suitable electron emitting materials [7, 8]. Its work function is about 4.3 eV [9] but, by doping and using electrostatic gating, a W-value as low as ~1 eV was achieved [10]. Moreover, because of the two-dimensional structure of graphene (one-atom thickness), electrons in graphene are at its boundary with the vacuum while in a 3D structure (bulk) material only the electrons near the surface are at such a boundary. In an electron emitting process, the electrons after passing the boundary can be emitted immediately, therefore graphene with whole electrons at its boundary act as an excellent electron emitter material [11].

This work presents experimental results on the lowering of the work function of graphene oxide (GO) via a chemical doping approach with cesium carbonate ($Cs_2CO_3$). Although it represents a first step and the manufactured samples are still far from the final application, the work is mainly focused on producing low-W graphene coating for LWT applications. For that reason, we used a conductor (aluminum and copper) as substrates for the doped GO layer. Such a configuration, i.e. a conductive substrate coated with doped GO, is new as compared to previous works on LWTs that suggested the use of a coating made of the C12A7:e- electride [12]. Section 2 justifies the selection of GO and Cs as dopant and explains the preparation of the doped GO and the coating process. As shown in Sec. 3, the samples were diagnosed via energy dispersive X-ray spectroscopy (EDS), Field Emission Scanning Electron microscopy (FE-SEM), four-point probe, and ultraviolet photoelectron spectroscopy (UPS). The results show that doping GO with $Cs_2CO_3$ (GO/Cs) can decrease the work function of GO from 4.6 eV to 3.09 eV. The conclusions of the work are summarized in Sec. 4.

## 2. Experimental details

Currently, chemical vapor deposition (CVD) is one of the most promising methods for growing high quality and large area graphene [13], which is faced with the high cost of production and the difficulty of transferring the grown graphene on target substrates [14, 15]. Chemically derived graphene oxide (GO) can be used as an alternative for CVD-grown graphene in order to produce graphene in a cost-effective and mass production manner. This member of graphene-family has a similar carbon backbone to graphene, while its surfaces are oxidized without disrupting the hexagonal graphene topology. Such properties indicate that GO can be a good choice to be combined with low-W materials in order to manufacture electron emitter devices. To obtain the low work function GO, there are different approaches [16] that among them, chemical doping arises as an effective and secure method. In this method, GO dopes with a low-W material such as alkali metal carbonates, which lead to a shift of the Fermi level of GO or the creation of a surface dipole that dramatically reduce the work function of GO [17, 18]. Low-W alkali metals (Li~2.93 eV, Na~2.36 eV, K~2.29 eV, Rb~2.26 eV, and Cs~1.95 eV) proved their applicability as electron injection materials in applications such as organic electronic devices including OLEDs and solar cells [17-19]. Due to the high solubility of the salt form of alkali metals in water and they spontaneously combination with carbon atoms on the graphene sheet, alkali metal carbonates are considered a simple and effective n-doping agent for lowering the work function of graphene sheets [20]. In most of the experimental approaches for lowering the work function of graphene as an electron emitter material, using CVD-grown graphene has been the most common method. Nevertheless, in order to coat the kilometer-long LWT with a low work function material, using CVD approach is not appropriate, while using the GO in the form of solution and using the spray coating technique for depositing the low work function material on a long conductive tape is much more versatile. Besides the doped GO, we chose aluminum and copper for the substrates (foils of 50 microns of thickness). Such a decision was driven by the LWT applications, which need a conductor with a high conductivity-to-density ratio.

In this work, we have used $Cs_2CO_3$ to lower the work function of GO. GO was synthesized using a typical modified Hummer's method. For doping GO, first 4 ml of the GO solution (0.5 g/L concentration) was dispersed in 20 ml deionized water in an ultrasonic bath for 60 minutes. Then we add 1ml of the $Cs_2CO_3$ solution (1 mol) to the GO solution and stir the solution for 30 minutes. Although we will not show results, we found that a higher concentration of $Cs_2CO_3$ in GO solution leads to agglomeration of the alkali metal carbonate on the GO sheets, which affects negatively to the adhesion of the final material to the substrates and decreases the conductivity of the coating. To purify the GO/Cs solution, the samples were washed by deionized water and filtered through the polyvinylidene fluoride (PVDF) membrane (0.5 μm) several times. By investigating the pH of the GO/Cs solution after washing and filtration process and comparing with the GO solution, the neutralization of the GO/Cs sample was confirmed. The final GO/Cs sample was dried at 50 °C overnight. The low work function coatings were then fabricated by first dispersing the GO and GO/Cs samples in deionized water (0.1 mg/mL concentration) and spraying the GO/Cs and GO solutions (0.1 mg/mL concentration) onto clean Al and Cu substrates (2cm×3cm) by means of an airbrush. The head pressure was 40 psi and the distance between the nozzle and substrates around 15 cm. Lastly, the obtained films were dried at 50 °C overnight in order to remove residual solvent. The GO and GO/Cs coatings were characterized via energy dispersive X-ray spectroscopy (EDS), Field Emission Scanning Electron microscopy (FE-SEM) and ultraviolet photoelectron spectroscopy (UPS).

## 3. Results and discussion

### 3.1 EDS and FE-SEM

Field-emission scanning electron microscopy (FE-SEM) was obtained using LEO SUPRA 55 (Carl Zeiss AG, Germany), equipped with an energy dispersive X-ray spectroscopy (EDS) instrument EDAX GENESIS 2000 (EDAX Inc., USA). The EDS analysis (Table 1) was performed to reveal the presence of Cs in the GO sample. From the EDS data, it can be seen that the Cs atoms presented in GO/Cs sample are accompanied by O and C; therefore, the GO sheets are successfully doped with Cs atoms.

Table 1. EDS analysis of the atomic percentage of the elements in GO and GO/Cs samples.

| Sample | O (at %) | C (at %) | Cs (at %) | Sheet resistance (Ω/sq) |
|---|---|---|---|---|
| GO | 40.94 | 59.06 | ----- | $3.4\times10^3$ |
| GO/Cs | 38.41 | 58.90 | 2.69 | $1.3\times10^4$ |

FE-SEM images have been obtained to study the state of the GO sheet before and after Cs doping. The FE-SEM images correspond to Al foils coated with GO/Cs and GO and highlight the uniformity of the coated samples. As shown in panels (a) and (b) in Fig. 1, there is no apparent difference between the morphological structure of GO before and after doping with Cs atoms, thus revealing that the Cs does not agglomerate on the graphene sheets and it does not form particles. Therefore, it can be concluded that the GO sheets uniformly Cs-doped and it can be expecting that changing the different characteristic of GO after Cs-doping must be homogenous. Since the GO is flexible, the produced coating is also flexible, which is an interesting property for LWT applications (note that the tape tether should be packaged in a reel before deployment). Simple tests, based on a manual bending of the samples and a later visual inspection of the color of the coating, indicated that the adhesion of GO and GO/Cs to the substrate was good.

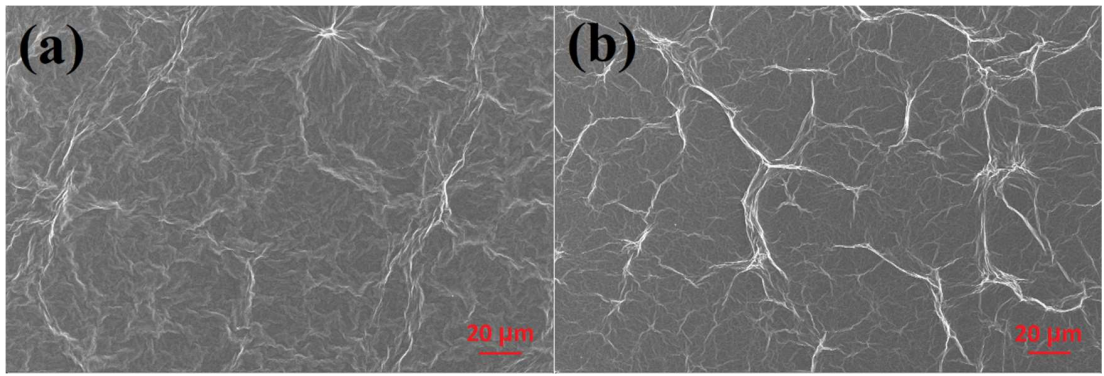

**Fig.1 FE-SEM images of (a) GO, and (b) GO/Cs samples.**

*3.2 Sheet resistance*

The sheet resistance of GO and GO/Cs samples was measured using a four-point probe system (ST-2258A) under ambient conditions from three different points on the sample surface. As it can be seen from the result (Table 1), due to the n-doping behavior of $Cs_2CO_3$, the sheet resistance of the GO sheets is increased compare to pristine GO sheets. Based on the reports, graphene is a p-type semiconducting material [21]; therefore, doping GO with an n-type dopant will increase the sheet resistance of GO, which indicates the charge transfer between the GO sheets and the dopant solution. Moreover, increasing the sheet resistance of GO point out the combination of metal dopant and carbon atoms.

*3.3 UPS analysis*

Ultraviolet photoelectron spectroscopy (UPS) was used to investigate the decrease in the GO work function by introducing $Cs_2CO_3$ among its structure. The UPS experiment was carried out in an Omicron Nanotechnology system with a base pressure of $2 \times 10^{-10}$ torr. UPS spectra were obtained using the He I irradiation (hv=21.2 eV). Samples were biased at -10 V during UPS measurements to avoid interference of the spectrometer threshold in the UPS spectra and observe the secondary electron edge. The work function was determined by subtracting the width of the photoelectron spectrum from the photon energy of the excitation light (hv). The former is calculated from the spectrum as the difference between the Fermi edge ($E_F$) and the inelastic high binding energy cutoff ($E_{cutoff}$). The work function of the sample W is then computed as:

$$W = hv - (E_F - E_{cutoff}) \qquad (1)$$

Fig. 2 shows the results of UPS measurement of GO and GO/Cs samples. As it is shown in Fig. 2a, the cut-off binding energy of GO is 16.6 eV. According to Eq. 1, the measured work function of GO is 4.6 eV, which is in agreement with previously reported values for GO (4.7-4.9 eV) [22, 23]. By doping GO with $Cs_2CO_3$ (GO/Cs), the cut-off binding energy shifts about 1.51 eV to higher binding energy and the work function of GO sheets reduced to 3.09 eV, which may facilitate the application of graphene as an electron source in different electronic, optic, and optoelectronic applications. Since the work function is the energy difference between the Fermi and the $E_{Cut-off}$ (Eq. 1) and no shift in the Fermi edge was observed after Cs-doping (Fig. 2b), the decreased work function can be attributed to the creation of surface dipoles.

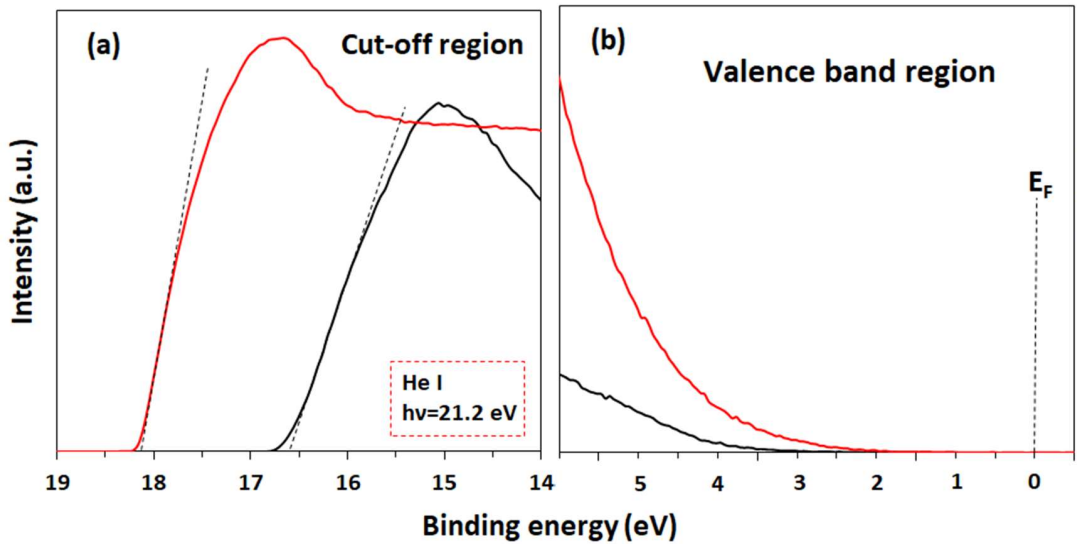

**Fig. 2 UPS spectrum of GO (black line) and GO/Cs (red line), (a) the cut-off region and (b) valance band spectra is shown magnified.**

## 4. Conclusions

This work proposes the manufacturing of Low Work-function Tether (LWT) samples by coating a conductive substrate (aluminum or copper) with graphene doped with Cs. Such a proposal was motivated by certain properties of the graphene, like its high electrical and thermal conductivity and the ability to withstand very high temperature, and the low work function of Cs, which is the Alkali metal with the lowest value (1.95eV). The proposed coating is novel with respect to previous works on LWTs, which suggested the use of the C12A7:e- electrode. One of the main advantages of the proposed coating is its low cost and easy implementation. First a GO solution was dispersed in deionized water in an ultrasonic bath and then we added $Cs_2CO_3$. After purification, the mixture was deposited in the conductive substrate with an air brush. Therefore, the method is scalable and could be used for coating km-length tapes, as needed for electrodynamic tether applications.

The experimental results via X-ray spectroscopy (EDS), Field Emission Scanning Electron microscopy (FE-SEM) and ultraviolet photoelectron spectroscopy (UPS) revealed important information about the coating. The EDS data showed that the GO sheets were successfully doped with Cs atoms and, from FE-SEM images before and after doping, we concluded that the Cs does not agglomerate on the graphene sheets and it does not form particles. The enhancement of the resistance of the doped GO sheets as compared to pristine GO sheets is consistent with the n-doping and p-doping behaviors of the $Cs_2CO_3$ and GO, respectively. The most important result of the work was provided by the UPS, which indicated that doping GO with Cs can lower the work function about 1.5eV, from 4.6 eV to 3.09 eV. Due to the sensitive dependence of the electron emission with the work function, such a result increases the applicability of GO as an electron emitter material considerably. However, it does not seem to be low enough for LWT applications and more analysis would be required. First optical analysis would be needed to determine the solar absorptance and infrared emittance of the samples, in order to compute the working temperature under solar illumination conditions. Such a temperature should be high enough to trigger the thermionic emission. Second, the determination of the photoelectric yield of the samples would be needed to evaluate the photoelectric current.

## Acknowledgements

This work was supported by Agencia Estatal de Investigación (Ministerio de Ciencia, Innovación y Universidades of Spain) under the project ESP2017-82092-ERC (AEI). SN work is supported by Comunidad de Madrid (Spain) under the Grant 2018/T2IND/11352. GSA work is supported by the Ministerio de Ciencia, Innovación y Universidades of Spain under the Grant RYC-2014-15357.

## References

1. NASA. *Graphene-Based Systems for Energy Storage*. 2018, June 09; Available from: https://ntrs.nasa.gov/search.jsp?R=20160013560.
2. Buffone, C., et al., *Capillary pressure in graphene oxide nanoporous membranes for enhanced heat transport in Loop Heat Pipes for aeronautics.* Exp. Therm Fluid Sci., 2016. **78**: p. 147-152.
3. Matloff, G.L., *Graphene, the ultimate interstellar solar sail material.* Journal of the British Interplanetary Society, 2012. **65**: p. 378-381.
4. Sanmartin, J., M. Martínez-Sánchez, and E. Ahedo, *Bare wire anodes for electrodynamic tethers.* J. Propul. Power, 1993. **9**(3): p. 353-360.
5. Williams, J.D., J.R. Sanmartin, and L.P. Rand, *Low work-function coating for an entirely propellantless bare electrodynamic tether.* IEEE Transactions on Plasma Science, 2012. **40**(5): p. 1441-1445.
6. Sanchez-Arriaga, G. and X. Chen, *Modeling and performance of electrodynamic low-work-function tethers with photoemission effects.* J Propul Power, 2017. **34**(1): p. 213-220.
7. Starodub, E., N.C. Bartelt, and K.F. McCarty, *Viable thermionic emission from graphene-covered metals.* Appl. Phys. Lett., 2012. **100**(18): p. 181604.
8. Wei, X., Y. Bando, and D. Golberg, *Electron emission from individual graphene nanoribbons driven by internal electric field.* ACS nano, 2011. **6**(1): p. 705-711.

9. Hibino, H., et al., *Dependence of electronic properties of epitaxial few-layer graphene on the number of layers investigated by photoelectron emission microscopy.* Physical Review B, 2009. **79**(12): p. 125437.
10. Yuan, H., et al., *Engineering ultra-low work function of graphene.* Nano Lett., 2015. **15**(10): p. 6475-6480.
11. Wu, Z.S., et al., *Field emission of single-layer graphene films prepared by electrophoretic deposition.* Adv. Mater., 2009. **21**(17): p. 1756-1760.
12. Toda, Y., et al., *Intense thermal field electron emission from room-temperature stable electride.* Appl. Phys. Lett., 2005. **87**(25): p. 254103.
13. Naghdi, S., K.Y. Rhee, and S.J. Park, *A catalytic, catalyst-free, and roll-to-roll production of graphene via chemical vapor deposition: Low temperature growth.* Carbon, 2018. **127**: p. 1-12.
14. Ye, D., et al., *Highly efficient electron field emission from graphene oxide sheets supported by nickel nanotip arrays.* Nano Lett., 2012. **12**(3): p. 1265-1268.
15. Naghdi, S., K.Y. Rhee, and S.J. Park, *Transfer-Free chemical vapor deposition of graphene on silicon substrate at atmospheric pressure: A sacrificial catalyst.* Thin Solid Films, 2018. **657**: p. 55-60.
16. Naghdi, S., G. Sanchez-Arriaga, and K.Y. Rhee, *Tuning the work function of graphene toward application as anode and cathode.* arXiv preprint arXiv:1905.06594, 2019.
17. Chen, Y., et al., *Graphene oxide-based carbon interconnecting layer for polymer tandem solar cells.* Nano Lett., 2014. **14**(3): p. 1467-1471.
18. Liu, J., et al., *Hole and electron extraction layers based on graphene oxide derivatives for high-performance bulk heterojunction solar cells.* Adv. Mater., 2012. **24**(17): p. 2228-2233.
19. Chang, J.-H., et al., *Solution-processed transparent blue organic light-emitting diodes with graphene as the top cathode.* Scientific reports, 2015. **5**: p. 9693.
20. Kwon, K.C., et al., *Work-function decrease of graphene sheet using alkali metal carbonates.* The Journal of Physical Chemistry C, 2012. **116**(50): p. 26586-26591.
21. Park, J., et al., *Work-function engineering of graphene electrodes by self-assembled monolayers for high-performance organic field-effect transistors.* The Journal of Physical Chemistry Letters, 2011. **2**(8): p. 841-845.
22. Ji, S., et al., *Work function engineering of graphene oxide via covalent functionalization for organic field-effect transistors.* Appl. Surf. Sci., 2017. **419**: p. 252-258.
23. Li, S.-S., et al., *Solution-processable graphene oxide as an efficient hole transport layer in polymer solar cells.* ACS nano, 2010. **4**(6): p. 3169-3174.